\documentclass[prb,reprint,twocolumn,citeautoscript,superscriptaddress,noeprint]{revtex4-2}
\usepackage{amsmath}
\usepackage{bm}
\usepackage{graphicx,color}
\usepackage{stix}
\usepackage{subfigure}
\usepackage{array}
\usepackage[usenames,dvipsnames]{xcolor}

\newcommand*{\veck}{{\mathbf{k}}}

\newcommand*{\gw}{{\textit{G}\textsubscript{0}\textit{W}\textsubscript{0}}}
\newcommand*{\GW}{\textit{GW}}
\newcommand*{\CC}{{\rm CC}}

\newcommand*{\full}{{\rm full}}

\definecolor{myblue}{rgb}{0,0,1}
\usepackage[breaklinks=true,colorlinks=true,linkcolor=myblue,urlcolor=myblue,citecolor=myblue]{hyperref}

\begin{document}
\title{Periodic Coupled-Cluster Green's Function for Photoemission Spectra of Realistic Solids}
\author{Katelyn Laughon}
\affiliation{Department of Chemistry, Yale University, New Haven, CT 06520}
\author{Jason M. Yu}
\affiliation{Department of Chemistry, University of California, Irvine, Irvine, CA 92697}
\author{Tianyu Zhu}
\email{tianyu.zhu@yale.edu}
\affiliation{Department of Chemistry, Yale University, New Haven, CT 06520}

\begin{abstract}
We present an efficient implementation of coupled-cluster Green's function (CCGF) method for simulating photoemission spectra of periodic systems. We formulate the periodic CCGF approach with Brillouin zone sampling in Gaussian basis at the coupled-cluster singles and doubles (CCSD) level. To enable CCGF calculations of realistic solids, we propose an active-space self-energy correction scheme by combining CCGF with cheaper many-body perturbation theory (\textit{GW}) and implement the model order reduction (MOR) frequency interpolation technique. We find that the active-space self-energy correction and MOR techniques significantly reduce the computational cost of CCGF while maintaining the high accuracy. We apply the developed CCGF approaches to compute spectral properties and band structure of silicon (Si) and zinc oxide (ZnO) crystals using triple-$\zeta$ Gaussian basis and medium-size $\mathbf{k}$-point sampling, and find good agreement with experimental measurements.
\end{abstract}

\maketitle

Accurate first-principles simulation of spectral properties is key to understanding and designing solid-state materials for energy, catalysis, and quantum technologies. Density functional theory (DFT)~\cite{Kohn1965} has been the workhorse for calculating band structure of solids due to its low cost, although it suffers from systematic errors and Kohn-Sham orbital energies do not formally describe the quasiparticle energies~\cite{Perdew1982,Perdew1983,Perdew2017}. Correlated Green's function methods provide a formal route to computing photoemission spectra beyond DFT~\cite{Hedin1965,VonNiessen1984,Peng2021,Hirata2015a,Hirata2017,Rusakov2016a,Banerjee2019,Backhouse2021}. One of the most successful Green's function approaches for this task is the many-body perturbation theory, or \textit{GW}~\cite{Hybertsen1986,Golze2019a,Shishkin2006,Deslippe2012,Zhu2021a,Ren2021}. Owing to proper treatment of dielectric screening, the \textit{GW} theory in its one-shot formulation predicts accurate band gaps of weakly-correlated semiconductors and insulators. On the other hand, no \textit{GW} formulation (e.g., self-consistency, DFT starting point, vertex correction) is known to be consistently reliable across weakly and strongly correlated materials. To achieve quantitative description of charged excitations beyond \textit{GW}, one promising framework is the Green's function embedding such as dynamical mean-field theory (DMFT)~\cite{Georges1992,Kotliar2006,Zhu2020,Zhu2021c} and self-energy embedding theory (SEET)~\cite{Rusakov2019b,Iskakov2020}, but other approximations must be invoked, which require careful treatment. 

Hence, it is necessary to develop higher-order \textit{ab initio} Green's function methods for periodic systems. Recently, the coupled-cluster (CC) theory has been extended to compute ground-state and excited-state properties of realistic solids and shows great promise in simulating both weakly (e.g., silicon) and strongly (e.g., nickel oxide) correlated materials~\cite{McClain2017,Gruber2018,Gao2020,Wang2020,Gallo2021a}. Meanwhile, molecular coupled-cluster Green's function (CCGF) implementations have been developed for studying photoelectron spectra of molecules and models~\cite{Nooijen1992,Nooijen1993,Bhaskaran-Nair2016,Peng2018a,Peng2019a,McClain2016} and solving impurity problems in \textit{ab initio} DMFT calculations~\cite{Zhu2019,Zhu2020,Zhu2021c,Shee2019,Shee2022}. However, efficient periodic CCGF implementation capable of simulating photoemission spectra and band structure of realistic materials is not yet available due to high computational cost~\cite{Furukawa2018a}. In this work, we fill this gap by developing accelerated periodic CCGF approach in Gaussian basis with Brillouin zone sampling. 

We start with a description of molecular CCGF theory~\cite{Bhaskaran-Nair2016,Peng2018a,Zhu2019}. The one-particle Green's function of a given system $G(\omega)=G^+(\omega)+G^-(\omega)$ in frequency (energy) domain is defined as:
\begin{subequations}
\begin{align}
 G^+_{pq}(\omega) &=  \langle \Psi_0 | {a}_p  
    \left[\omega - (\hat{H}-E)+i\eta\right]^{-1}  {a}^\dag_q | \Psi_0\rangle , \\
  G^-_{pq}(\omega) &= \langle \Psi_0 | {a}^\dag_q  
    \left[\omega + (\hat{H}-E)-i\eta\right]^{-1}  {a}_p | \Psi_0\rangle ,
\end{align}
\label{eq:exactgf}%
\end{subequations}
where $G^+(\omega)$ and $G^-(\omega)$ are addition (EA) and removal (IP) parts of Green's function, $|\Psi_0\rangle$ is the ground-state wave function, $\hat{H}$ is the Hamiltonian, $E$ is the ground-state energy, and $\eta$ is a small broadening factor. $a_p$ and $a_q^\dag$ are annihilation and creation operators on orbitals $p$ and $q$. In coupled-cluster theory, the CC ground-state wave function is parameterized as
\begin{equation}
  |\Psi_0\rangle = e^{\hat{T}} |\Phi_0\rangle ,
\label{eq:cc}
\end{equation}
with $\hat{T}$ as the cluster excitation operator and $|\Phi_0\rangle$ as the Hartree-Fock determinant. In this work, we truncate the $\hat{T}$ operator at the singles and doubles level (i.e., CCSD). The CC bra state is parameterized differently as:
\begin{equation}
  \langle \Psi_0 | = \langle \Phi_0 | (1+\hat{\Lambda}) e^{-\hat{T}} ,
\label{eq:ccbra}
\end{equation}
where $\hat{\Lambda}$ is the de-excitation operator. Inserting Eq.~\ref{eq:cc} and Eq.~\ref{eq:ccbra} into Eq.~\ref{eq:exactgf}, one arrives at the CCGF equations:
\begin{subequations}
\begin{align}
 G^+_{pq}(\omega) &=  \langle \Phi_0 | (1+\hat{\Lambda}) \bar{a}_p  
    \left[\omega - (\bar{H}-E)+i\eta\right]^{-1}  \bar{a}^\dag_q | \Phi_0\rangle , \\
  G^-_{pq}(\omega) &= \langle \Phi_0 | (1+\hat{\Lambda}) \bar{a}^\dag_q  
    \left[\omega + (\bar{H}-E)-i\eta\right]^{-1}  \bar{a}_p | \Phi_0\rangle ,
\end{align}
\label{eq:ccgf}%
\end{subequations}
where similarity transformed operators are defined as:
\begin{align}
\bar{a}_p &= e^{-\hat{T}} \hat{a}_p e^{\hat{T}}, \notag \\
\bar{a}_p^\dagger &= e^{-\hat{T}} \hat{a}_p^\dagger e^{\hat{T}}, \notag \\
\bar{H} &= e^{-\hat{T}} \hat{H} e^{\hat{T}} .
\end{align}

To efficiently solve Eq.~\ref{eq:ccgf}, we define vectors $Y_q(\omega)$ and $X_p(\omega)$:
\begin{subequations}
\begin{align}
 \left[\omega - (\bar{H}-E)+i\eta\right] Y_q(\omega) | \Phi_0\rangle &=   
     \bar{a}^\dag_q | \Phi_0\rangle , \\
 \left[\omega + (\bar{H}-E)-i\eta\right] X_p(\omega) | \Phi_0\rangle &=   
      \bar{a}_p | \Phi_0\rangle ,
\end{align}
\label{eq:xvec}%
\end{subequations}
so that Eq.~\ref{eq:ccgf} becomes
\begin{subequations}
\begin{align}
 G^+_{pq}(\omega) &=  \langle \Phi_0 | (1+\hat{\Lambda}) \bar{a}_p Y_q(\omega) | \Phi_0\rangle , \\
  G^-_{pq}(\omega) &= \langle \Phi_0 | (1+\hat{\Lambda}) \bar{a}^\dag_q X_p(\omega) | \Phi_0\rangle .
\end{align}
\end{subequations}
To solve the set of linear equations in Eq.~\ref{eq:xvec}, $Y_q(\omega)$ and $X_p(\omega)$ are parameterized in the EOM-CCSD (equation-of-motion CCSD) approximation~\cite{Stanton1993,Krylov2008}:
\begin{subequations}
\begin{align}
 Y_q(\omega) &= \sum_a y^a(q, \omega) a_a^\dagger + \sum_{i, a<b} y_i^{ab} (q, \omega) a_a^\dagger a_b^\dagger a_i ,  \\
  X_p(\omega) &= \sum_i x_i(p, \omega) a_i + \sum_{i<j, a} x_{ij}^{a} (p, \omega) a_a^\dagger a_i a_j  ,
\end{align}
\label{eq:xvec2}%
\end{subequations}
where $Y_q(\omega)$ contains one-particle ($1p$) and two-particle one-hole ($2p1h$) terms and $X_p(\omega)$ contains $1h$ and $2h1p$ terms. We use the sparse linear equation solver GCROT(m,k)~\cite{DeSturler1999} as implemented in SciPy~\cite{2020SciPy-NMeth} to solve Eqs.~\ref{eq:xvec} and \ref{eq:xvec2}.

For periodic systems, we work explicitly with $\veck$-point (Brillouin zone) sampling in the reciprocal space. The $\veck$-point CCGF equation is
\begin{subequations}
\begin{align}
 G^+_{pq}(\veck, \omega) &=  \langle \Phi_0 | (1+\hat{\Lambda}) \bar{a}_{p\veck}  
    \left[\omega - (\bar{H}-E)+i\eta\right]^{-1}  \bar{a}^\dag_{q\veck} | \Phi_0\rangle , \\
  G^-_{pq}(\veck, \omega) &= \langle \Phi_0 | (1+\hat{\Lambda}) \bar{a}^\dag_{q\veck}  
    \left[\omega + (\bar{H}-E)-i\eta\right]^{-1}  \bar{a}_{p\veck} | \Phi_0\rangle .
\end{align}
\label{eq:kccgf}%
\end{subequations}
From the Green's function, one obtains momentum-dependent spectral function
\begin{equation}
A(\veck, \omega)= -\frac{1}{\pi} \mathrm{Im}G(\veck, \omega),
\end{equation}
whose trace defines density of states (DOS).

We implemented $\veck$-point CCGF approach in PySCF quantum chemistry software package~\cite{Sun2020b} using a hybrid MPI+OpenMP parallelization scheme. To avoid solving $\veck$-point $\Lambda$ equations, we approximate the $\Lambda$ amplitudes as the complex conjugate of $T$ amplitudes. We evaluated the accuracy of this approximation against an exact supercell (molecular) CCGF calculation on the diamond crystal and found excellent agreement (Fig. S1). 

\onecolumngrid

\begin{figure*}[htb]
    \centering
     \begin{subfigure}
         \centering
         \includegraphics[width=0.36\textwidth]{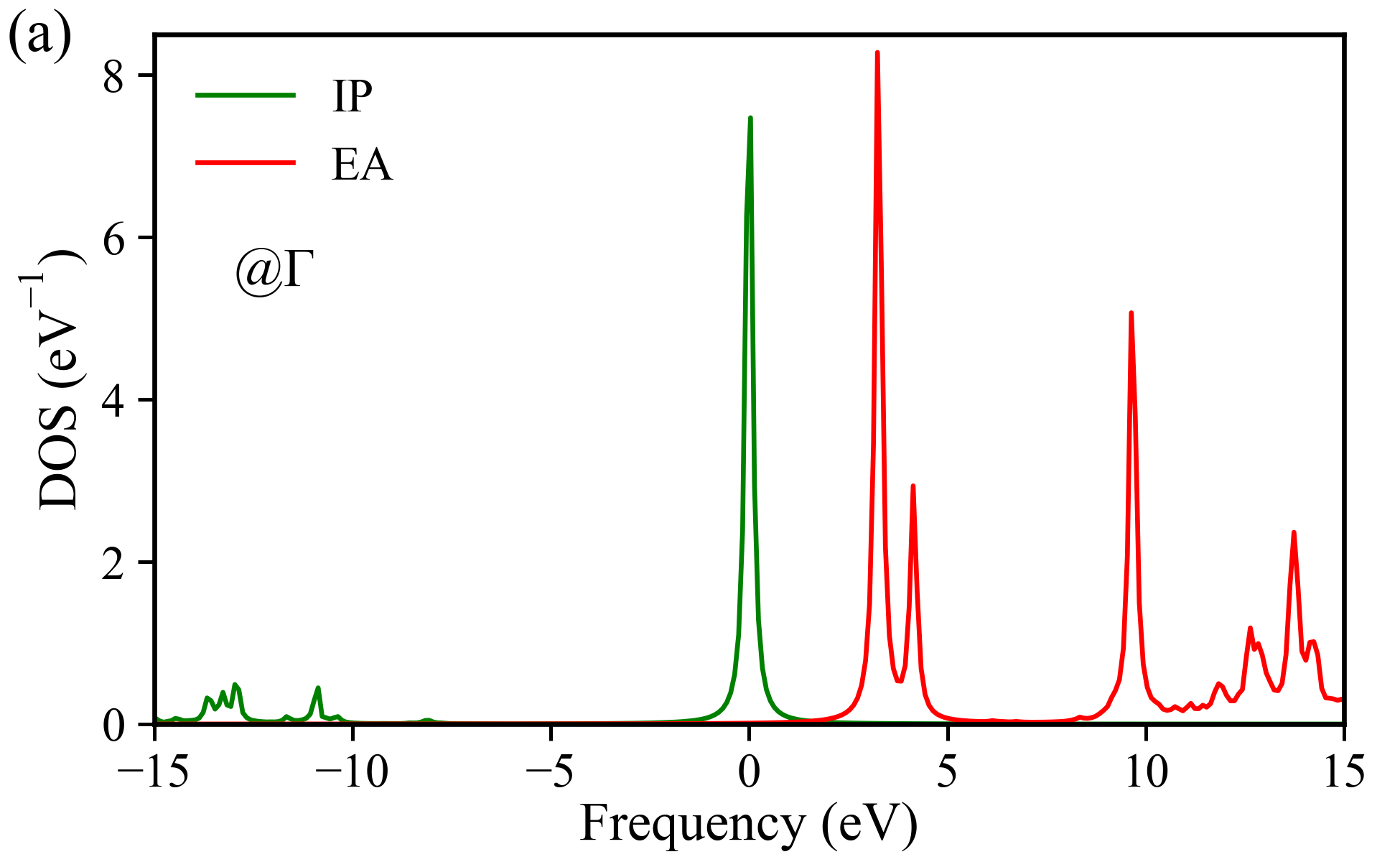}
     \end{subfigure}
     \hspace{1em}%
     \begin{subfigure}
         \centering
         \includegraphics[width=0.36\textwidth]{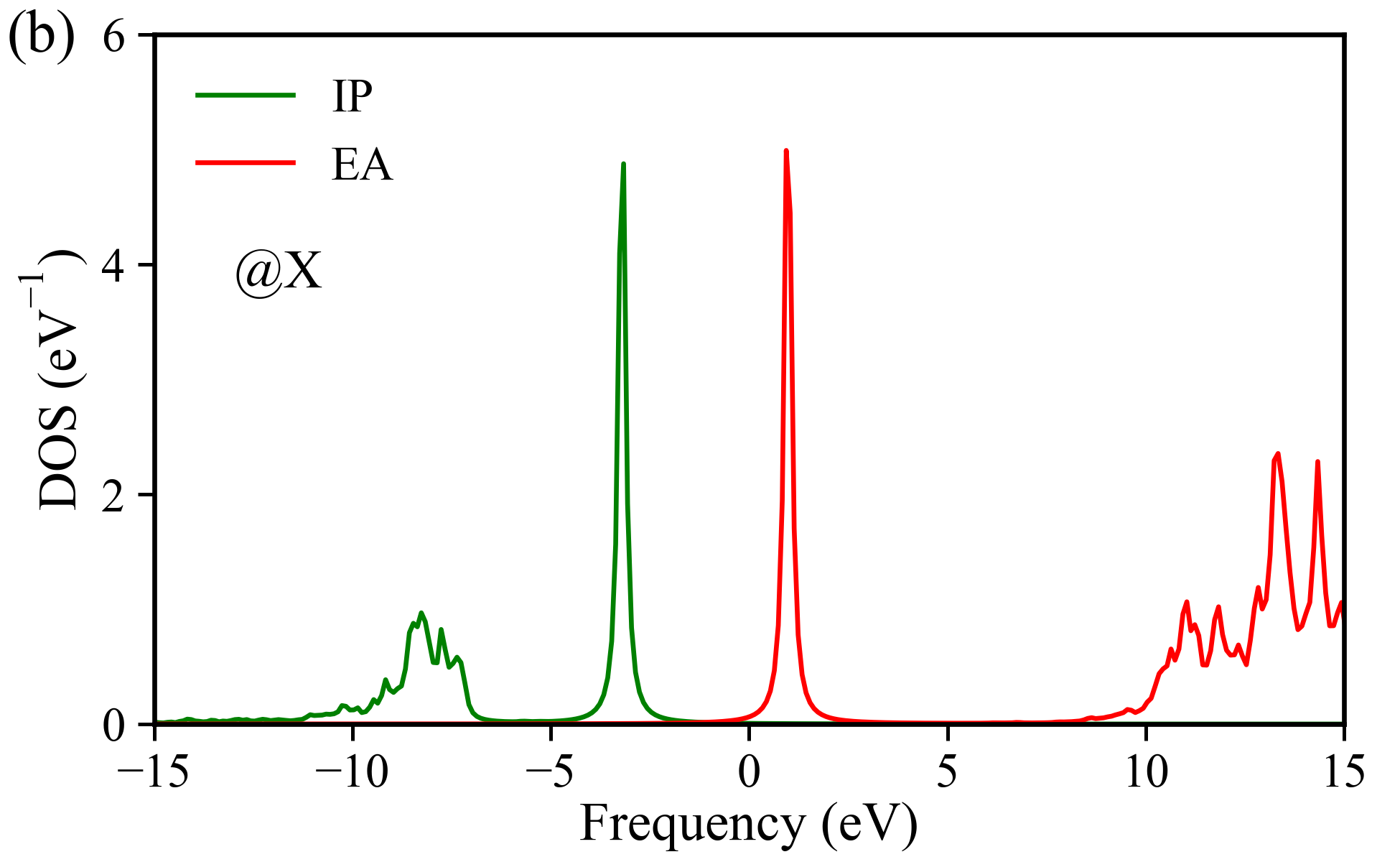}
     \end{subfigure}
     \vskip\baselineskip
     \begin{subfigure}
         \centering
         \includegraphics[width=0.36\textwidth]{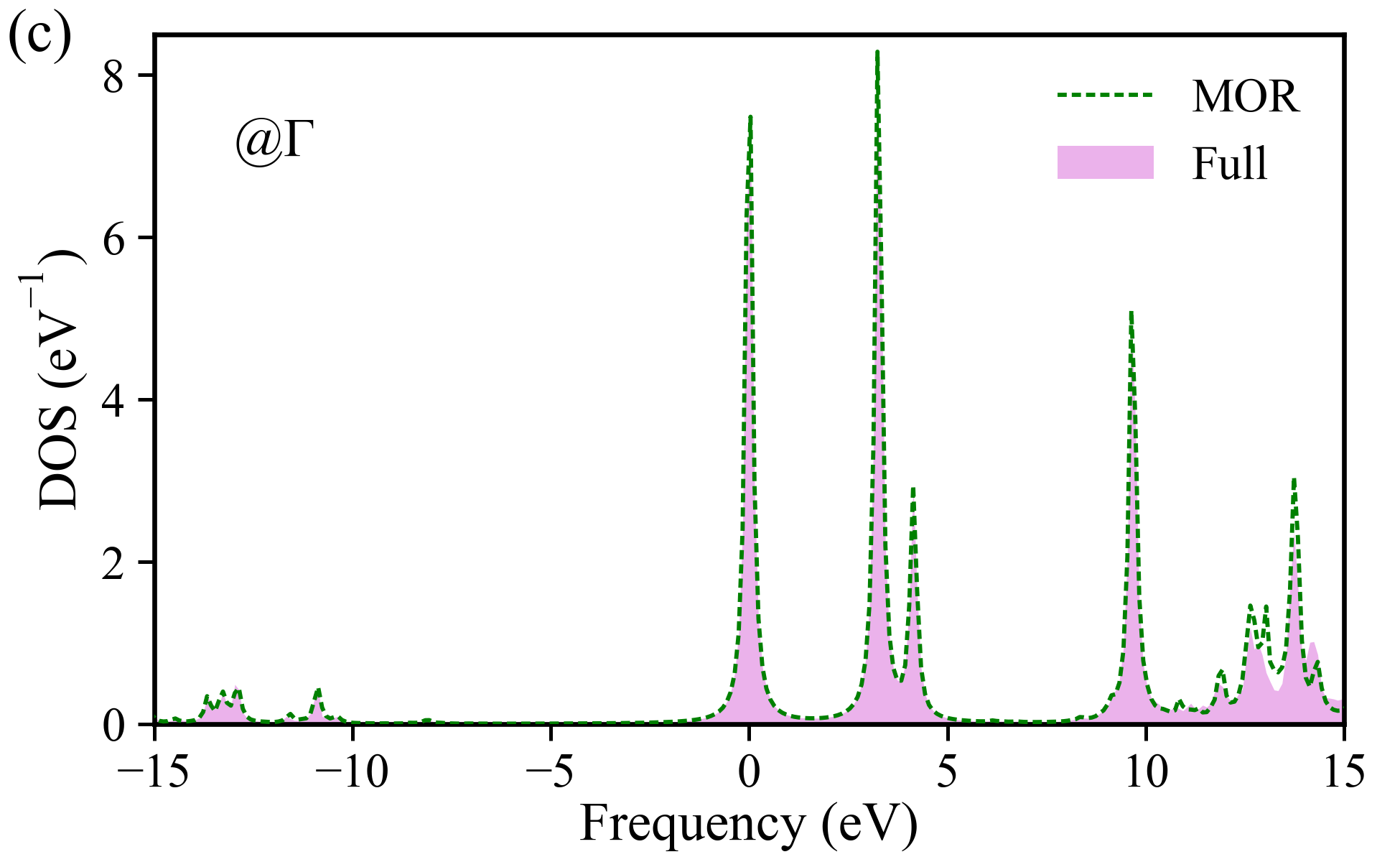}
     \end{subfigure}
     \hspace{1em}%
     \begin{subfigure}
         \centering
         \includegraphics[width=0.36\textwidth]{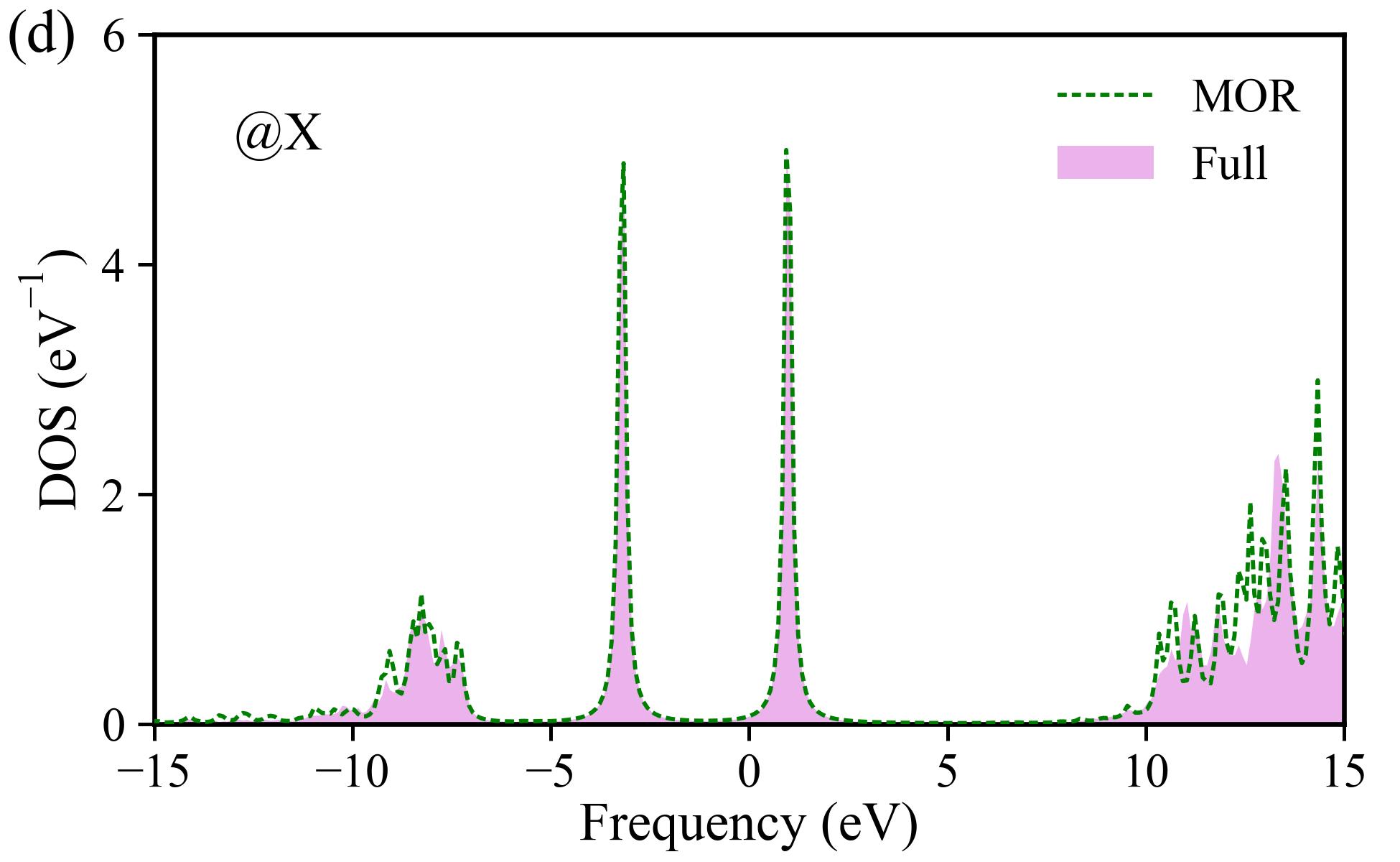}
     \end{subfigure}
    \caption{Density of states of silicon crystal computed by periodic CCGF. GTH-DZVP basis set and $2\times2\times2$ $\veck$-mesh were used. DOS is shifted so that valence band maximum (at $\Gamma$) is centered at 0 eV. A broadening factor of 0.1 eV is used. (a) DOS at $\Gamma$. (b) DOS at X. (c) MOR-CCGF against full CCGF at $\Gamma$. (d) MOR-CCGF against full CCGF at X.}
    \label{fig:CCGFMOR}
\end{figure*}
\twocolumngrid

In Fig.~\ref{fig:CCGFMOR}(a) and \ref{fig:CCGFMOR}(b), we show momentum-dependent density of states (DOS) of Si computed by $\veck$-point CCGF method with GTH-HF pseudopotential and GTH-DZVP basis set~\cite{Hartwigsen1998,Vandevondele2005} as well as $2\times 2 \times 2$ $\veck$-point sampling. Because our periodic CCGF approach is formulated on real-frequency axis, it is capable of computing photoemission spectra of valence, core, and high-virtual bands, which involve quasiparticle and satellite (e.g., [-15, -10] eV and [10, 15] eV in Fig.~\ref{fig:CCGFMOR}(a)) peaks.

Although it is now possible to perform small $\veck$-point CCGF calculations, the high computational scaling of periodic CCGF prohibits its application to more complex materials. At each $\veck$-point, frequency $\omega$, and Green's function column/row (for EA/IP), the scaling of the EA part is $\mathcal{O}(N_\veck^3 N_o N_v^4)$ and the scaling of the IP part is $\mathcal{O}(N_\veck^3 N_o^3 N_v^2)$. $N_\veck, N_o, N_v$ are the number of sampled $\veck$-points, occupied orbitals, and virtual orbitals (per unit cell, respectively). The overall cost for computing $G^+(\veck, \omega)$ and $G^-(\veck, \omega)$ matrices for all $\veck$-points are thus $\mathcal{O}(N_\omega N_\veck^4 N N_o N_v^4)$ and $\mathcal{O}(N_\omega N_\veck^4 N N_o^3 N_v^2)$, with $N=N_o+N_v$ and $N_\omega$ as the sampled frequency points. In the rest of this Letter, we describe acceleration techniques to reduce $N_\omega$ and $N$.

\textit{Model order reduction}. To obtain high resolution in photoemission spectra, one usually needs to perform CCGF calculations on hundreds of frequency points. This large prefactor can be significantly lowered by the model order reduction (MOR) method, which is a technique for reducing computational complexity of mathematical models and has been successfully applied to compute X-ray absorption spectra and CCGF for molecules~\cite{VanBeeumen2017a,Peng2019a}. We refer the readers to Ref.~\cite{Peng2019a} for details of the MOR-CCGF implementation. Briefly speaking, in MOR-CCGF, one computes the full CCGF (Eq.~\ref{eq:xvec}) on a small set of selected frequency points (e.g., $N_\omega^\mathrm{MOR} \approx 10-20$), then uses solved $X(\omega)$ or $Y(\omega)$ to construct a subspace. The original effective Hamiltonian ($\bar{H}$) is then projected onto the subspace to form a much smaller model with dimension of $N_\omega^\mathrm{MOR} \times N_\omega^\mathrm{MOR}$. One finally solves CCGF equations on all $N_\omega$ frequency points ($N_\omega \approx 200-400$) using this reduced model, which has negligible cost. Thus, MOR is a frequency interpolation (sometimes extrapolation) technique that decreases the prefactor from $N_\omega$ to $N_\omega^\mathrm{MOR}$.

We implemented the MOR technique within our periodic CCGF code and tested the accuracy of MOR-CCGF on Si. In Fig.~\ref{fig:CCGFMOR}(c) and ~\ref{fig:CCGFMOR}(d), we chose $N_\omega^\mathrm{MOR}=20$ (equally distributed on [-15,0] eV for IP and [0,15] eV for EA) and $N_\omega=321$ respectively (meaning the cost of MOR-CCGF is $1/16$ of full CCGF). MOR-CCGF reproduces the main quasiparticle peaks around Fermi level perfectly compared to full CCGF calculations. Even the satellite peaks at [-15, -5] eV and [10, 15] eV are captured accurately. If one focuses only on the valence peaks, we found that $N_\omega^\mathrm{MOR}=8$ is enough to yield highly accurate CCGF DOS (see Fig. S2).

\textit{Active-space self-energy correction.} Following the idea of frozen natural orbital coupled-cluster theory~\cite{Taube2005,Taube2008,Landau2010a}, we develop an active-space approach to reduce the number of orbitals in periodic CCGF calculations. Specifically, we show an efficient combination of CCGF and \textit{GW} methods through a self-energy correction scheme. 

We first perform a \textit{GW} calculation on the system using Gaussian-based \gw~method developed by one of the authors (T.Z.)~\cite{Zhu2021a}. Throughout this Letter, we employ the one-shot \gw@HF method that scales as $\mathcal{O}(N_\veck^2 N^4)$. From the \gw@HF calculation, we obtain the \textit{GW} density matrix $\gamma^\textit{GW}(\veck, \omega)$ using a linearized \textit{GW} density matrix formalism~\cite{Bruneval2019} which guarantees conserving particle number. The \textit{GW} density matrix is then diagonalized to derive a set of natural orbitals:
\begin{equation}
 \gamma^\GW(\veck) V(\veck) = V(\veck) n(\veck) ,
\label{eq:gwno}
\end{equation}
where $V(\veck)$ is the natural orbital (NO) coefficient and $n(\veck)$ is the occupation number of NOs. We then select $M$ most partially occupied orbitals to form an active space and perform periodic CCGF calculation. Because $M<N$, the computational cost of CCGF is approximately decreased from $\mathcal{O}(N_\omega N_\veck^4 N^6)$ to $\mathcal{O}(N_\omega N_\veck^4 M^6)$. We note that we keep active space the same size for all $\veck$-points to simplify the implementation. 

\onecolumngrid

\begin{figure*}[htb]
    \centering
     \begin{subfigure}
         \centering
         \includegraphics[width=0.33\textwidth]{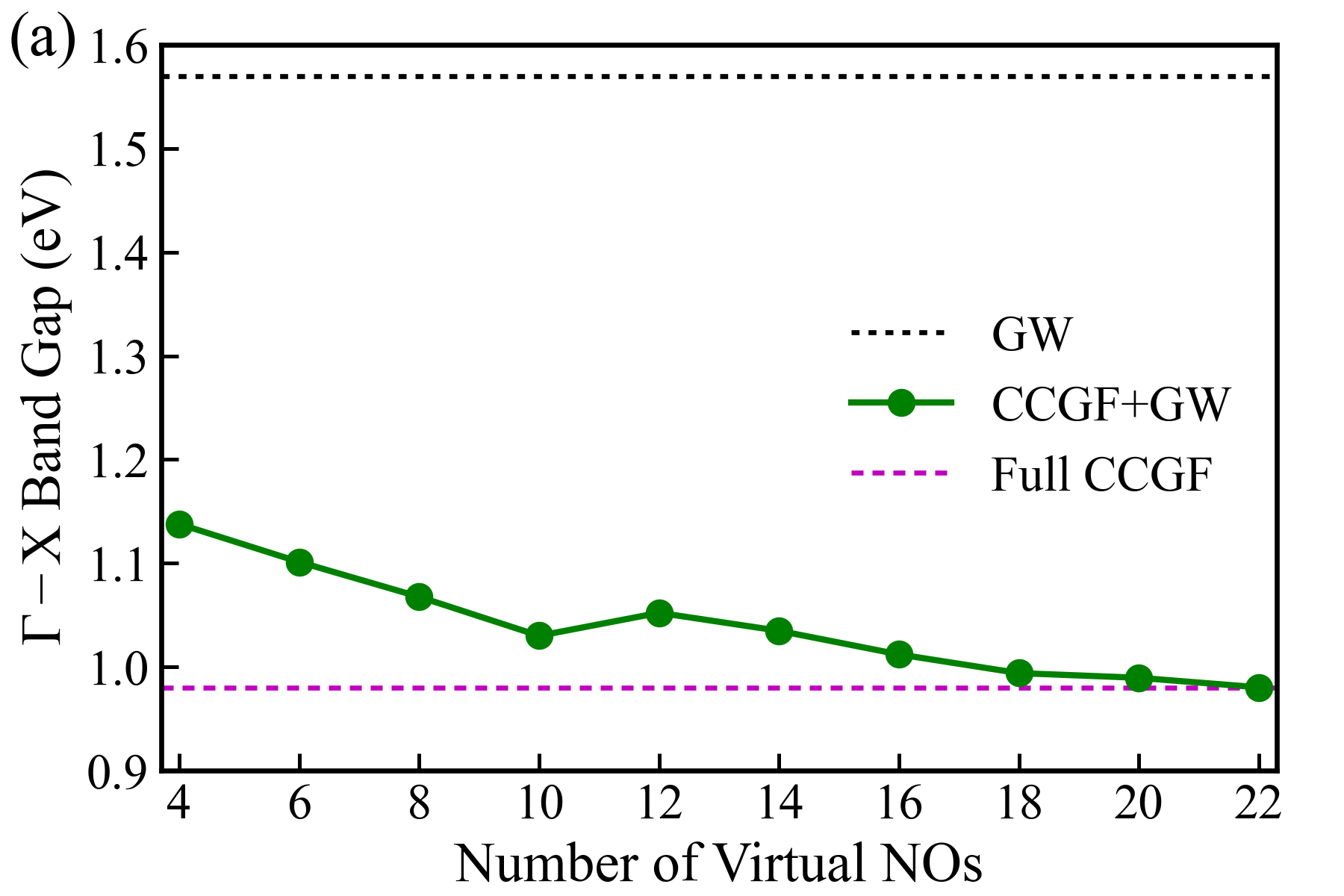}
     \end{subfigure}
     \hspace{1em}%
     \begin{subfigure}
         \centering
         \includegraphics[width=0.36\textwidth]{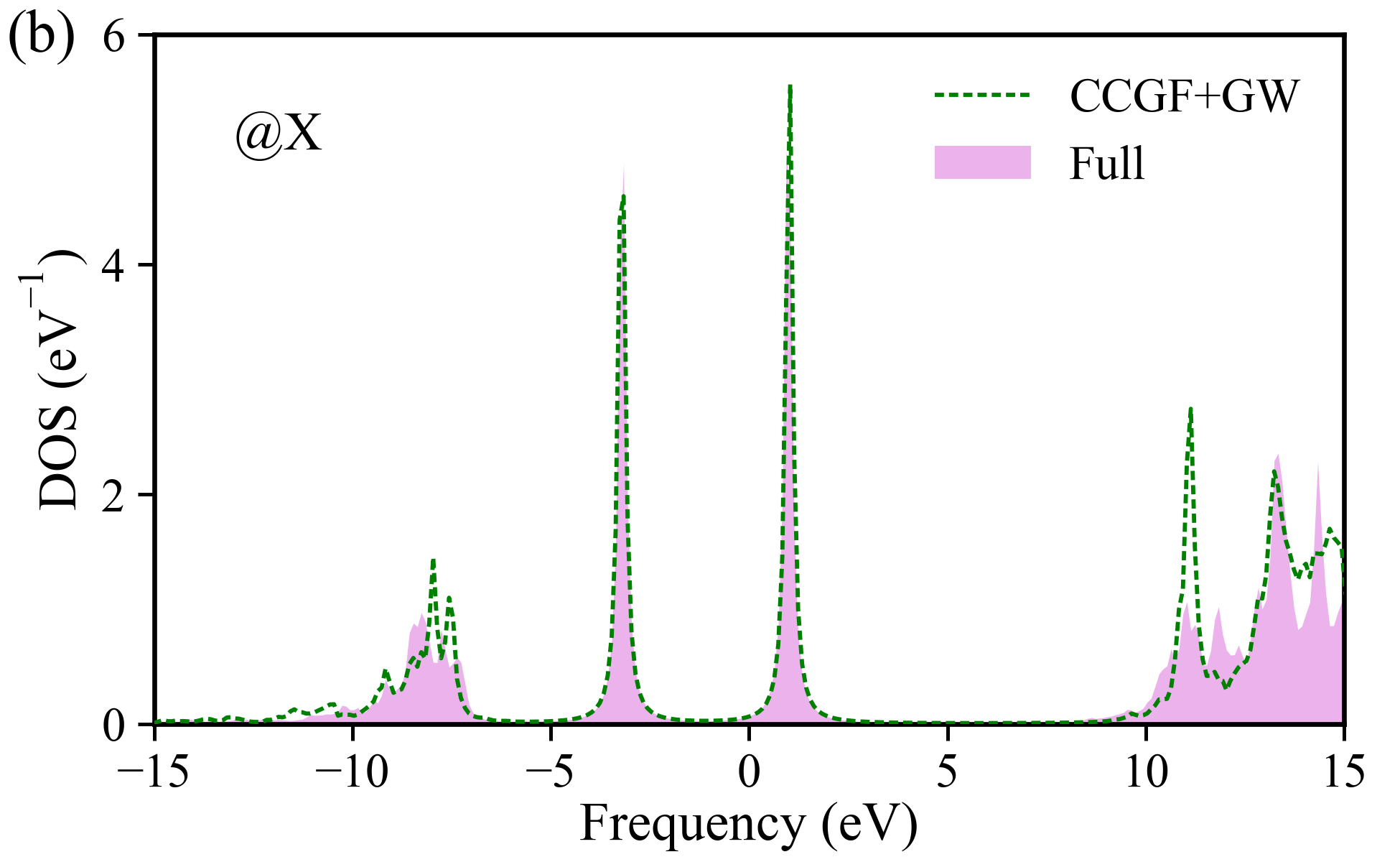}
     \end{subfigure}
     \vskip\baselineskip
     \begin{subfigure}
         \centering
         \includegraphics[width=0.33\textwidth]{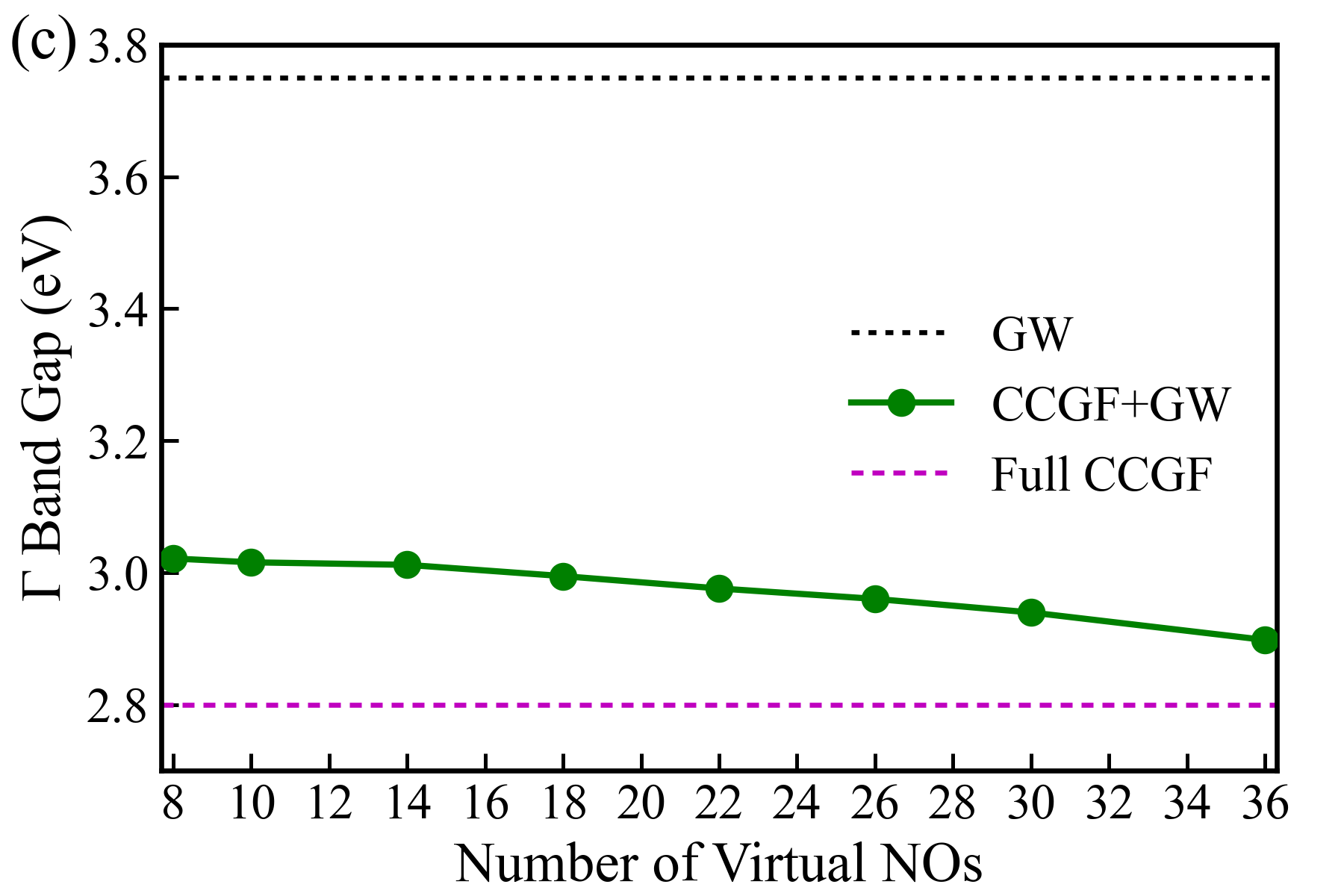}
     \end{subfigure}
     \hspace{1em}%
     \begin{subfigure}
         \centering
         \includegraphics[width=0.36\textwidth]{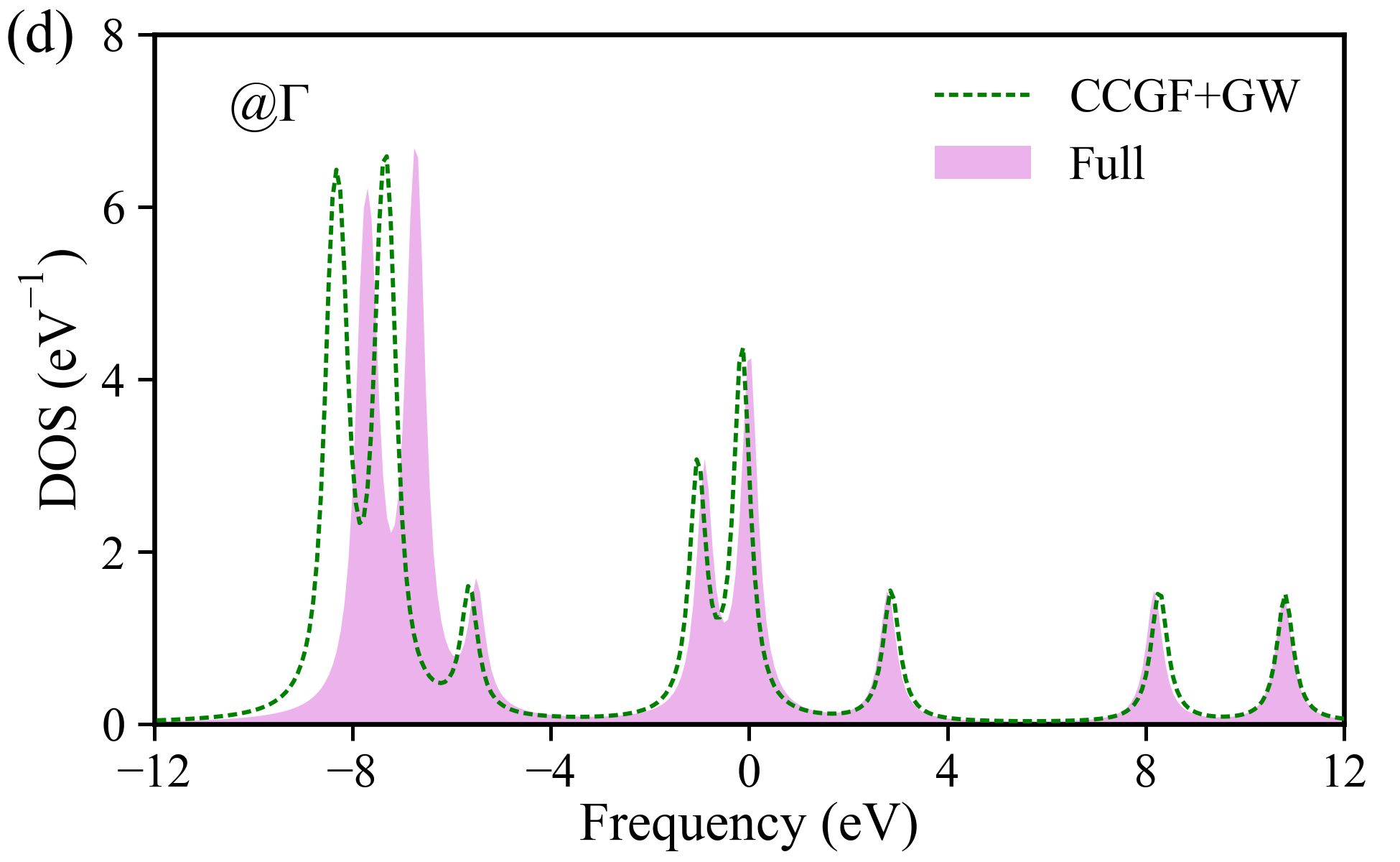}
     \end{subfigure}
    \caption{Periodic CCGF results with active-space self-energy correction. (a) $\Gamma-$X band gap of Si computed by CCGF+\GW~as a function of virtual NO number per unit cell. Full CCGF and \gw@HF results are also included. (b) DOS for Si at X computed by (8e, 14o) active-space CCGF+\GW~and full CCGF. $\eta=0.1$ eV. (c) Band gap of wurtzite ZnO computed by CCGF+\GW~compared against full CCGF and \gw@HF. MOR ($N_\omega^\mathrm{MOR}=8$) is employed for full CCGF and CCGF+\GW~calculations. (d) DOS for ZnO at $\Gamma$ computed by (16e, 18o) active-space CCGF+\GW~and full CCGF. $\eta=0.2$ eV.} 
    \label{fig:CAS}
\end{figure*}
\twocolumngrid

However, this frozen natural orbital scheme does not work well for excited states and truncating the NO basis from $N$ to $M$ results in a loss of accuracy. To remedy the error, we define a self-energy correction term within the complete active space (CAS):
\begin{equation}
 \Sigma_\mathrm{CAS}(\veck, \omega) = G_{\GW, \mathrm{CAS}}^{-1} (\veck, \omega) - G_{\CC, \mathrm{CAS}}^{-1}(\veck, \omega) ,
\label{eq:gwact}
\end{equation}
and transform it back to the full molecular orbital (MO) space 
\begin{equation}
 \Sigma_\full(\veck, \omega) =   V(\veck) \Sigma_\mathrm{CAS}(\veck, \omega) V(\veck)^\dag .
\end{equation}
Then the CC+\textit{GW} Green's function is computed through Dyson's equation:
\begin{equation}
 G_{\CC+\GW}(\veck, \omega) = [G_{\GW, \full}^{-1} (\veck, \omega) - \Sigma_\full(\veck, \omega)]^{-1} ,
\label{eq:ccgw}
\end{equation}
where the \textit{GW} Green's function ($G_{\GW, \full}$) for the full system is computed when deriving the natural orbital basis. We note that this scheme can also be applied using other low-level theories, such as MP2 or self-consistent \textit{GW}~\cite{Shishkin2007,VanSchilfgaarde2006a}. 

We benchmarked the accuracy of CCGF+\GW~on Si and wurtzite zinc oxide (ZnO). We used GTH-DZVP basis set and $2\times2\times2$ $\veck$-point sampling for Si and def2-SV(P) basis set~\cite{Weigend2005a} and $2\times2\times1$ $\veck$-point sampling for ZnO. In Fig.~\ref{fig:CAS}(a), we tested the $\Gamma-$X band gap of Si using different number of virtual NOs in the active space, while all (4) occupied orbitals are included. As a comparison, the full CCGF and \gw@HF band gaps are 0.98 eV and 1.57 eV, which correspond to 22 virtual orbitals per unit cell. As the number of active virtual NOs increases, the CCGF+\GW~band gap quickly converges to the full CCGF result. We find that using only 4 virtual NOs per unit cell, the CCGF+\GW~band gap is 1.14 eV, only 0.16 eV larger than full CCGF gap. At 10 virtual NOs, the CCGF+\GW~band gap error is only 0.05 eV (or 5\%). We note that for DOS at $\Gamma$ point, the CCGF+GW~calculation with 10 virtual NOs takes 2 hours on 112 CPU cores, while the full CCGF calculation costs 19 hours using same resources.

The DOS for Si at X point is presented in Fig.~\ref{fig:CAS}(b) using an active space of (8e, 14o) (i.e., 8 electrons and 14 orbitals per unit cell). It is shown that not only the low-energy quasiparticle peaks are accurately described by CCGF+\GW, but also the deeper valence and higher virtual spectra are well captured.

In Fig.~\ref{fig:CAS}(c) and \ref{fig:CAS}(d), we present similar tests for ZnO. The full CCGF and \gw@HF band gap for ZnO are 2.80 eV and 3.75 eV, which correspond to 38 occupied and 38 virtual orbitals per unit cell. Fixing the number of active occupied orbitals to 8 in Fig.~\ref{fig:CAS}(c), reasonably accurate band gaps are produced by CCGF+\GW. For example, at 10 virtual NOs, the CCGF+\GW~band gap is 3.02 eV and has 8\% relative error compared to full CCGF. More test results on ZnO are available in the SI. The photoemission spectrum computed by (16e, 18o) CCGF+\GW~also shows good agreement with full CCGF in Fig.~\ref{fig:CAS}(d). The discrepancy in the [-10, -6] eV region is likely due to the small number of active occupied orbitals used in CCGF+\GW.

\begin{figure}[hbt]
\centering
\includegraphics{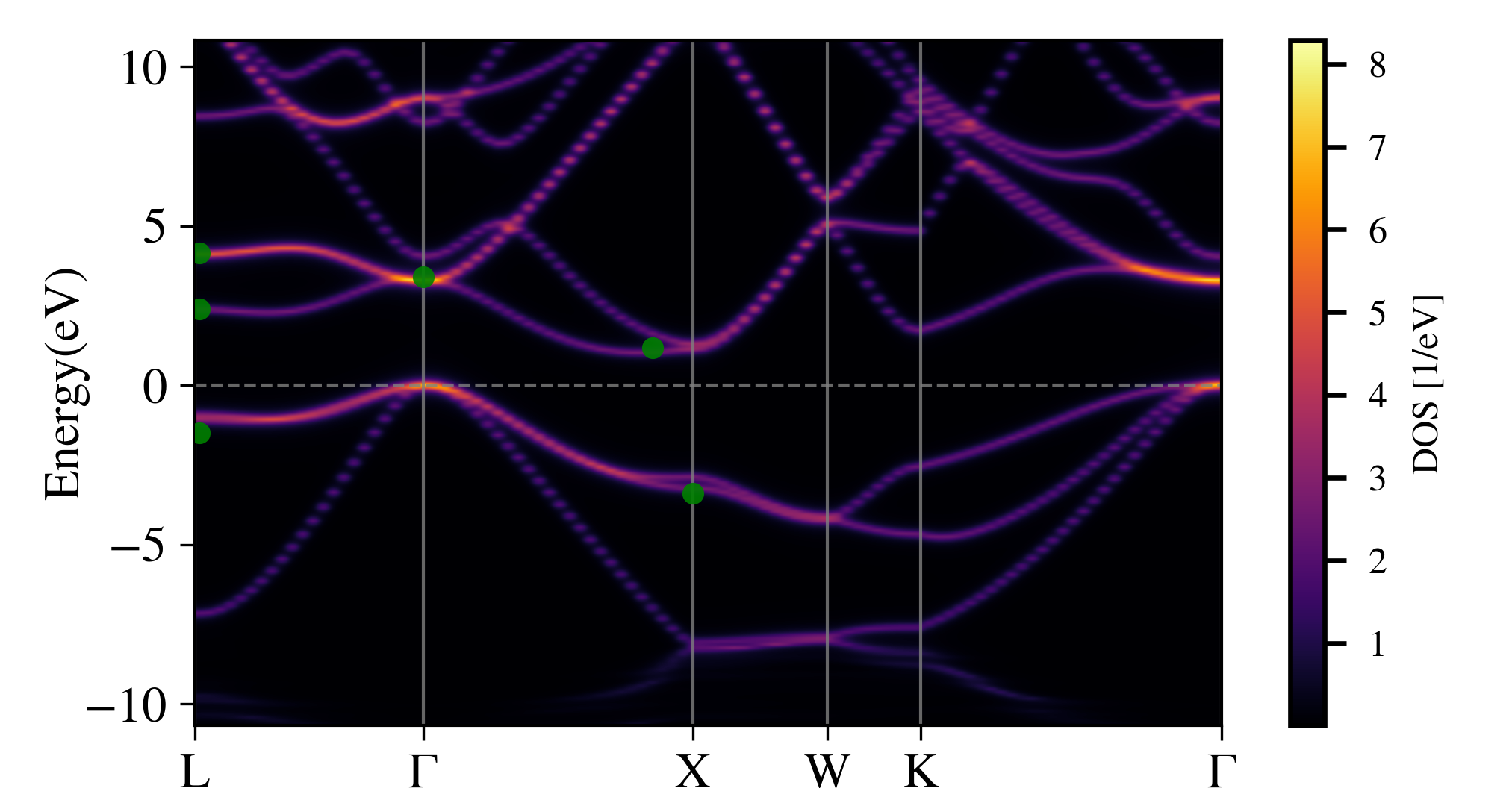}
\caption{Band structure of Si computed by MOR-CCGF+\GW~at $3\times3\times3$ $\veck$-point sampling. Experimental data~\cite{goldmann1989} are plotted in green circles.}
\label{fig:si}
\end{figure}

Combining both active-space self-energy correction and MOR techniques, we study Si and ZnO using higher-quality basis set and larger Brillouin zone sampling. For Si, we used the correlation-consistent GTH-cc-pVTZ basis set recently developed by Ye and Berkelbach~\cite{Ye2022} as well as GTH-HF pseudopotential. We applied MOR-CCGF+\GW~approach with an active space of (8e, 14o) per unit cell and $N_\omega^\mathrm{MOR}=8$. Using band interpolation technique based on intrinsic atomic orbital and projected atomic orbital (IAO+PAO)~\cite{Knizia2013d,Hamann2009a,Furukawa2018a}, we obtain band structure of Si from MOR-CCGF+\GW~calculation at $3\times3\times3$ $\veck$-mesh in Fig.~\ref{fig:si}. We find that the CCGF band structure is in excellent agreement with experimental photoemission data~\cite{goldmann1989}. 

To obtain Si band gap in the thermodynamic limit (TDL), we performed MOR-CCGF+\GW~calculations at $2\times2\times2$, $3\times3\times3$, and $4\times4\times4$ $\veck$-meshes, and conducted a finite size extrapolation of band gap with respect to $N_\veck^{-1/3}$. The results are summarized in Table~\ref{tab:bandgap}. Furthermore, we applied a correction term at $2\times2\times2$ $\veck$-mesh that accounts for the error introduced by using small active space:
\begin{equation}
    \Delta_\mathrm{CAS} = E_g(\mathrm{L}, 2\times2\times2) - E_g(2\times2\times2)
\end{equation}
where $E_g(\mathrm{L}, 2\times2\times2)$ corresponds to a (8e, 28o) MOR-CCGF+\GW~calculation and $E_g(2\times2\times2)$ refers to the (8e, 14o) calculation. The final estimated Si band gap is thus
\begin{align}
    E_g(\mathrm{TDL}+\Delta_\mathrm{CAS}) = E_g(\mathrm{TDL}) + \Delta_\mathrm{CAS}.
\end{align}
We find that $E_g(\mathrm{TDL})=1.01$ eV and $\Delta_\mathrm{CAS}=-0.03$ eV, leading to estimated CCGF band gap at 0.98 eV, which is 0.25 eV underestimated compared to the experimental band gap of 1.23 eV (taking zero-point renormalization effect into account). We note that our CCGF band gap is 0.21 eV smaller than EOM-CCSD result (1.19 eV) in Ref.~\cite{McClain2017}, which is mainly caused by the use of different basis sets (GTH-cc-pVTZ vs. GTH-TZVP).

\begin{table}[hbt]
	\centering
	\caption{Band gaps (eV) of Si and wurtzite ZnO computed by MOR-CCGF+\GW~at different $\veck$-point sampling. Finite size and CAS corrected (Extrap.+$\Delta_\mathrm{CAS}$) band gaps are also included. Experimental values are taken from Ref.~\cite{Ren2021} and corrected for the zero-point renormalization effect.}
	\label{tab:bandgap}
	\begin{tabular}{cccccc}
	\hline\hline
	 &  $2\times2\times2$ ~ &  $3\times3\times3$ ~ &  $4\times4\times4$ ~ & TDL+$\Delta_\mathrm{CAS}$ & Expt. \\
	\hline
Si   &   0.78  &  0.85   &  0.89 & 0.98  & 1.23 \\
    \hline
&  $2\times2\times1$ ~ &  $3\times3\times2$ ~ &  $4\times4\times3$ ~ & TDL+$\Delta_\mathrm{CAS}$ & Expt. \\
	\hline
ZnO  &  2.39 & 3.47 & 3.97 & 4.94 & 3.60  \\

    \hline\hline
	\end{tabular}
\end{table}

\begin{figure}[hbt]
\centering
\includegraphics{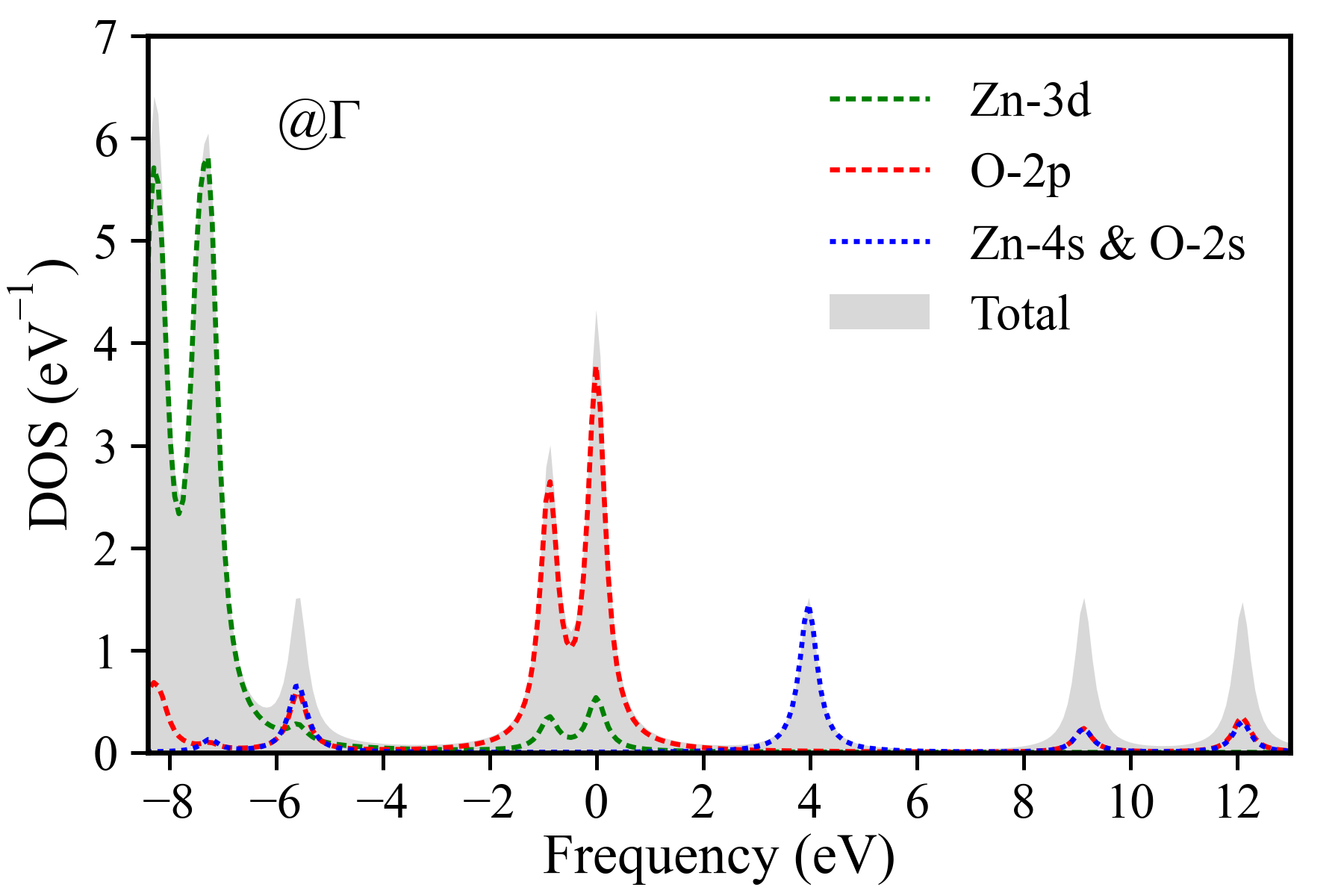}
\caption{Density of states of wurtzite ZnO at $\Gamma$ point computed by MOR-CCGF+\GW~with $4\times4\times3$ $\veck$-point sampling. $\eta=0.2$ eV.}
\label{fig:zno}
\end{figure}

We then report MOR-CCGF+\GW~results on wurtzite ZnO, using an active space of (16e, 18o) per unit cell and $N_\omega^\mathrm{MOR}=8$. We employed cc-pVTZ-PP basis set and pseudopotential~\cite{Figgen2005,Peterson2005a} for Zn and cc-pVTZ basis set~\cite{Dunning1989} for O. We present the DOS at $\Gamma$ point with $4\times4\times3$ $\veck$-point sampling in Fig.~\ref{fig:zno}, since wurtzite ZnO has direct band gap at $\Gamma$. By plotting the orbital-resolved DOS, we find that the valence band maximum (VBM) of ZnO is mainly contributed from O-$2p$ orbitals, with minor character of Zn-$3d$. On the other hand, the conduction band minimum (CBM) has dominant Zn-$4s$ and O-$2s$ components. 

We also find that the MOR-CCGF+\GW~band gap is 3.97 eV at $4\times4\times3$ $\veck$-mesh, only 0.37 eV larger than the experimental band gap of 3.60 eV. However, this good agreement is partially due to fortuitous error cancellation. The finite size extrapolated band gap of ZnO from $2\times2\times1$, $3\times3\times2$, $4\times4\times3$ $\veck$-point calculations is 5.18 eV. The error from using small active space is estimated according to 
\begin{equation}
    \Delta_\mathrm{CAS} = E_g(\mathrm{L}, 2\times2\times1) - E_g(2\times2\times1) ,
\end{equation} 
where $E_g(\mathrm{L}, 2\times2\times1)$ refers to a (52e, 76o) MOR-CCGF+\GW~calculation and $E_g(2\times2\times1)$ is the (16e, 18o) calculation. $\Delta_\mathrm{CAS}$ is computed to be $-$0.24 eV, thus our estimated CCGF band gap in the TDL is 4.94 eV, which is 1.34 eV overestimated than the experimental value. This study indicates CCGF at the EOM-CCSD level is not enough to produce quantitative accuracy in describing band gap of ZnO.

In conclusion, we developed efficient periodic coupled-cluster Green's function method and enabled simulating photoemission spectra of materials with high-quality basis set and realistic $\veck$-point sampling. We proposed and implemented active-space self-energy correction and MOR schemes, which significantly accelerate expensive periodic CCGF calculations. Periodic CCGF provides a higher-order Green's function tool than the commonly-used \GW~approximation, which is particularly attractive for benchmarking low-level theories and quantum embedding methods~\cite{Cui2020,Zhu2021c} on spectral properties of solids. 

\textit{Code availability.} The periodic CCGF code and examples are available at \href{https://github.com/ZhuGroup-Yale/kccgf}{https://github.com/ZhuGroup-Yale/kccgf}.

\begin{acknowledgments}
This work was supported by a start-up fund from Yale University. The National Science Foundation Graduate Research Fellowship Program (DGE-1745301) is acknowledged for support of J.M.Y. We thank Garnet Chan for helpful discussions and the Yale Center for Research Computing for supercomputing resources. This work also used the Extreme Science and Engineering Discovery Environment (XSEDE), which is supported by National Science Foundation grant number ACI-1548562. 
\end{acknowledgments}

%

\raggedbottom

\widetext
\clearpage
\begin{center}
\textbf{\large Supporting Information for: \\ Periodic Coupled-Cluster Green's Function for Photoemission Spectra of Realistic Solids}
\end{center}
\setcounter{equation}{0}
\setcounter{figure}{0}
\setcounter{table}{0}
\setcounter{page}{1}
\makeatletter
\renewcommand{\theequation}{S\arabic{equation}}
\renewcommand{\thefigure}{S\arabic{figure}}
\renewcommand{\thetable}{S\arabic{table}}
\renewcommand{\bibnumfmt}[1]{[S#1]}
\renewcommand{\citenumfont}[1]{S#1}

\section{Validation of $\veck$-point CCGF code}
\begin{figure}[hbt]
\centering
\includegraphics{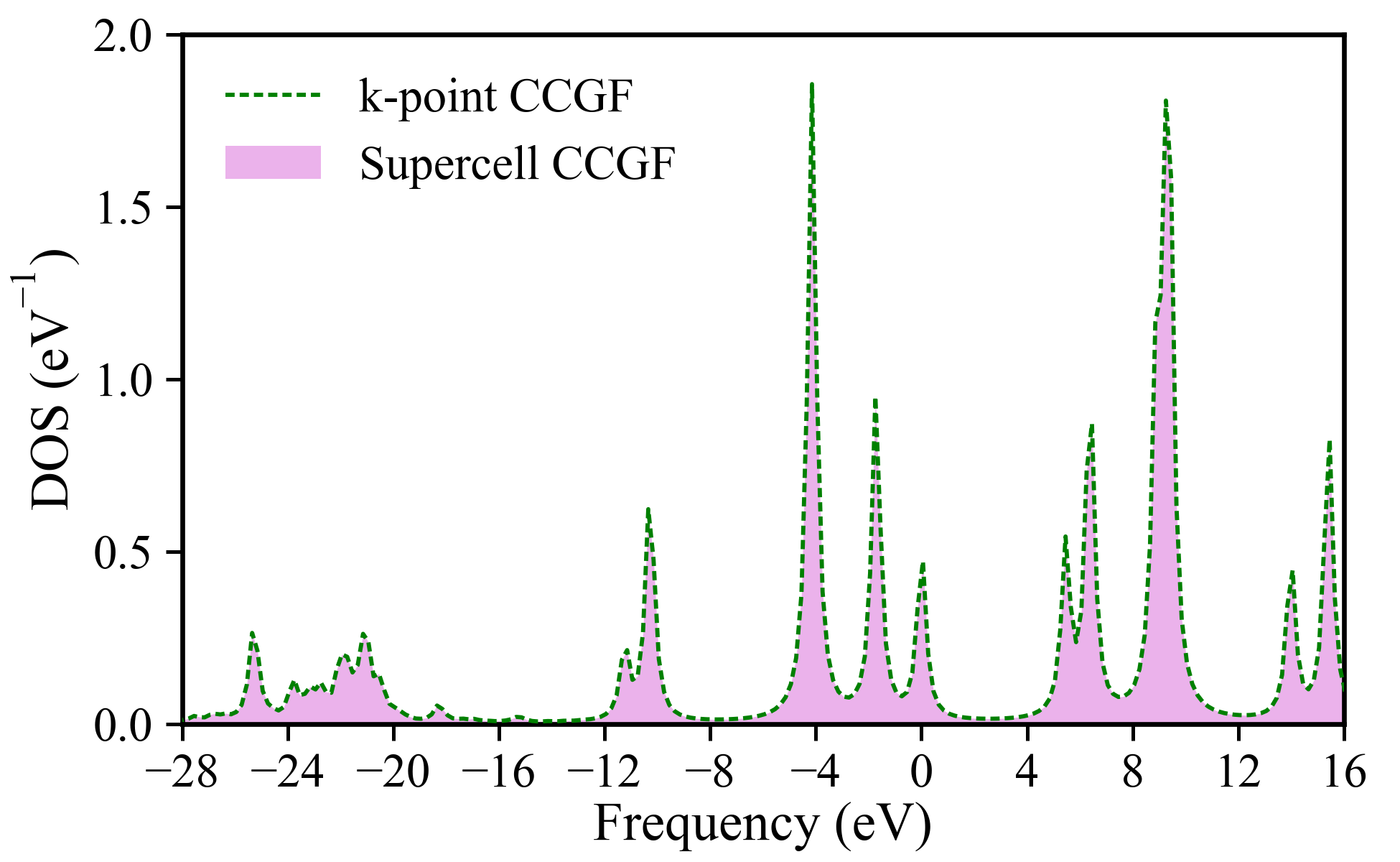}
\caption{Local DOS on diamond crystal computed by periodic CCGF. $\eta=0.2$ eV.}
\label{fig:supercell}
\end{figure}

We tested the $\veck$-point CCGF implementation with $\Lambda$ amplitudes approximated by complex conjugate of $T$ amplitudes against a supercell (molecular) CCGF implementation without approximation on diamond crystal. We used $3\times1\times1$ $\veck$-point sampling and GTH-DZVP basis sets. As shown in Fig.~\ref{fig:supercell}, $\veck$-point CCGF DOS is in excellent agreement with supercell CCGF DOS.

\section{MOR benchmark}
\begin{figure*}[htb]
    \centering
     \begin{subfigure}
         \centering
         \includegraphics[width=0.4\textwidth]{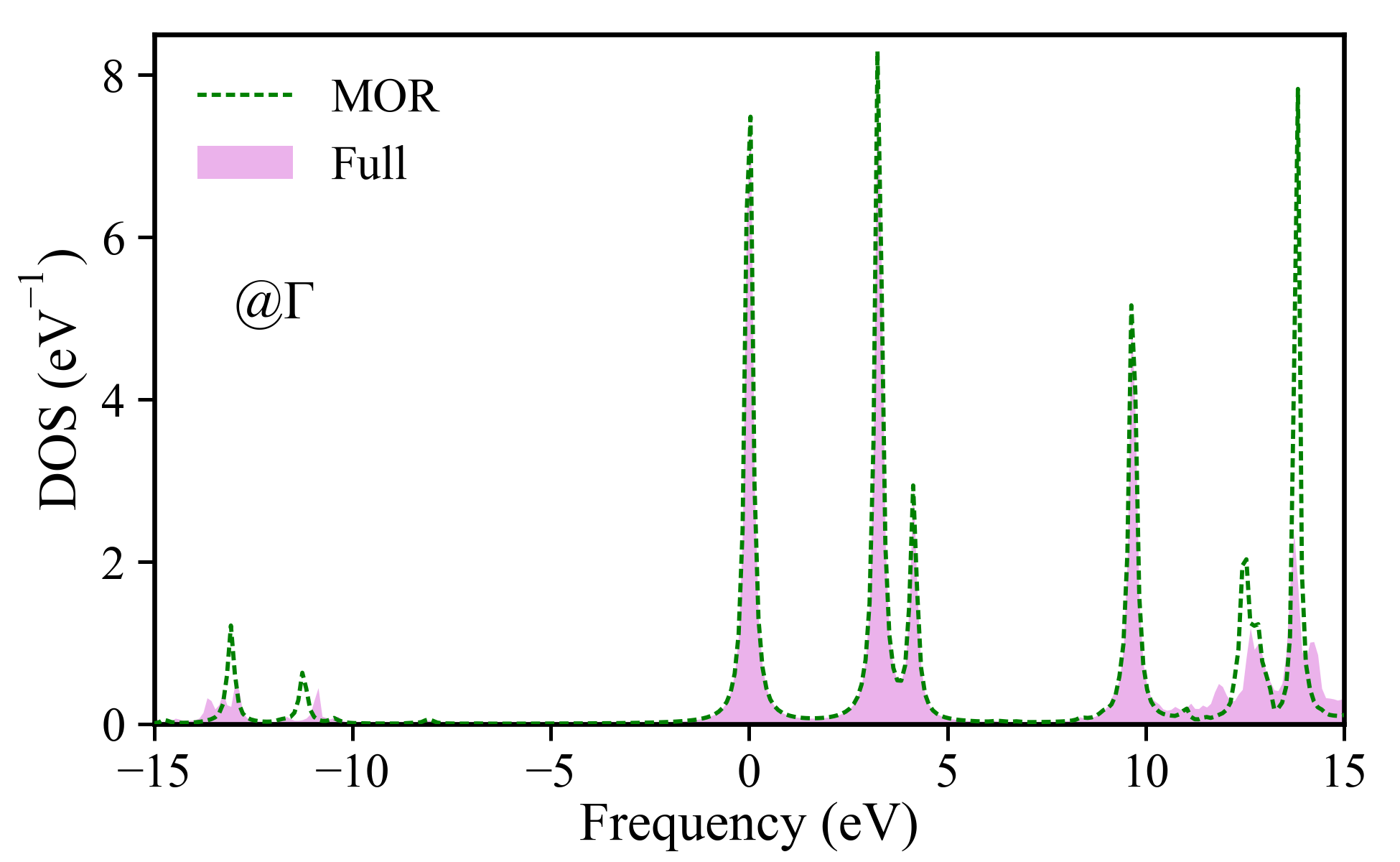}
     \end{subfigure}
     \hspace{1em}%
     \begin{subfigure}
         \centering
         \includegraphics[width=0.4\textwidth]{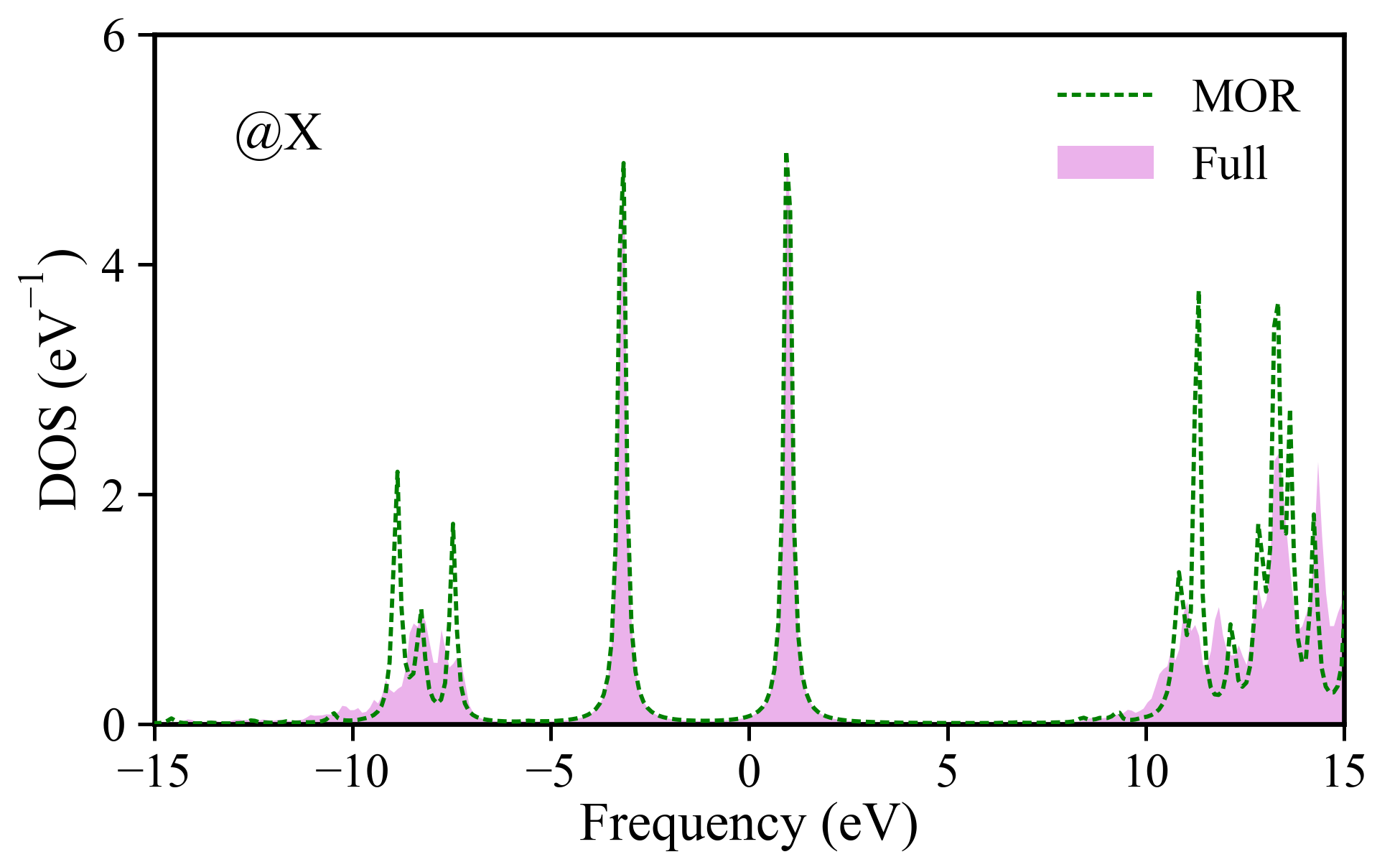}
     \end{subfigure}
    \caption{DOS of Si computed by MOR-CCGF and full CCGF. $N^\mathrm{MOR}_\omega=8$ and $\eta=0.1$ eV. Left: DOS at $\Gamma$. Right: DOS at X.} 
    \label{fig:MORSI}
\end{figure*}

In addition to $N^\mathrm{MOR}_\omega=20$, we also tested different number of MOR frequency points for Si (GTH-DZVP basis and $2\times2\times2$ $\veck$-mesh). As seen in Fig.~\ref{fig:MORSI}, we find that $N^\mathrm{MOR}_\omega=8$ is sufficient to obtain highly accurate CCGF quasiparticle peaks around the Fermi level, although the accuracy in the deep valence and high virtual regions is sacrificed.

\begin{figure}[hbt]
\centering
\includegraphics{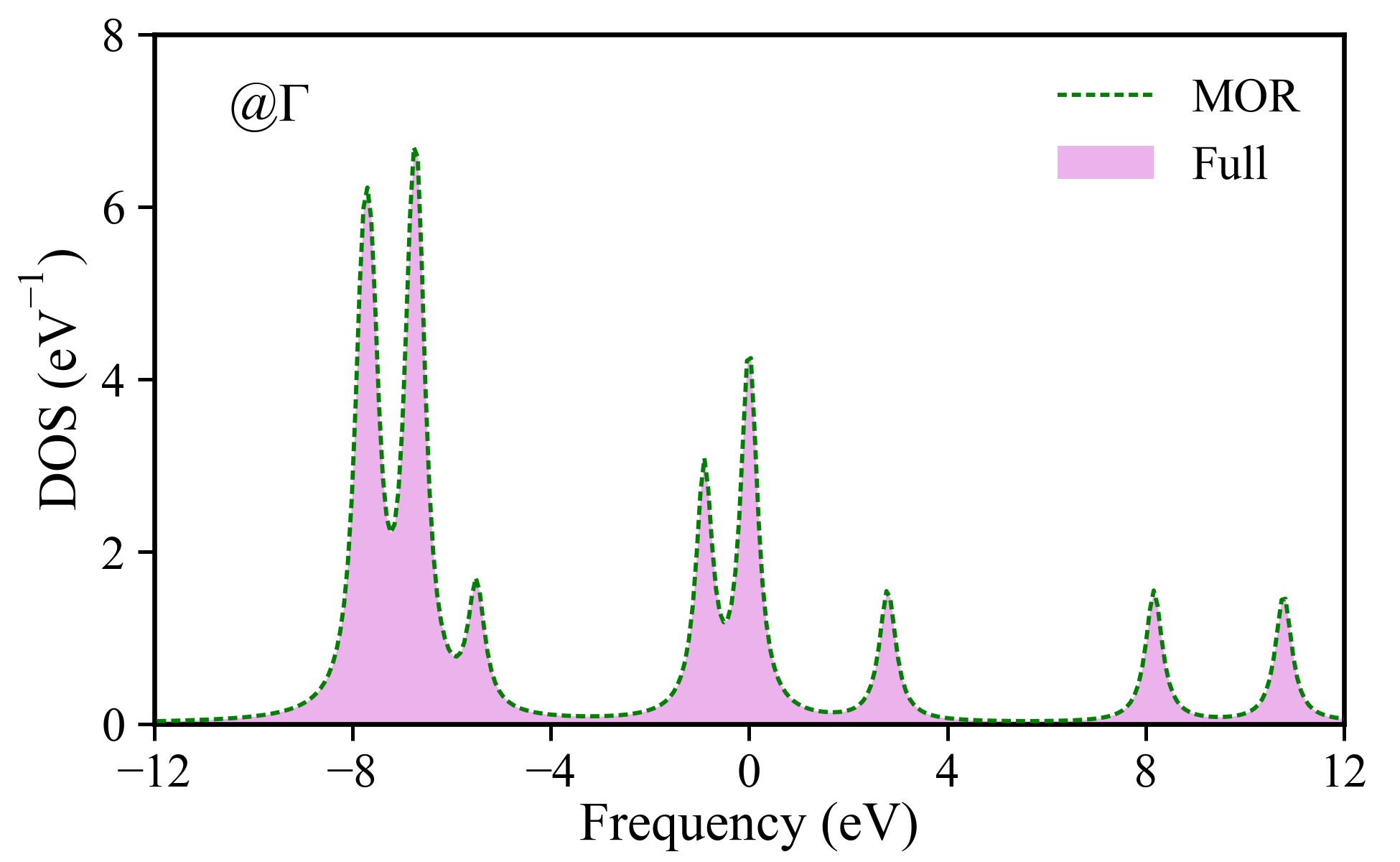}
\caption{DOS of wrutzite ZnO at $\Gamma$ point computed by MOR-CCGF and full CCGF. $N^\mathrm{MOR}_\omega=8$ and $\eta=0.2$ eV.}
\label{fig:MORzno}
\end{figure}

In Fig.~\ref{fig:MORzno}, we show a test of MOR-CCGF on wurtzite ZnO using def2-SV(P) basis set and $2\times2\times1$ $\veck$-point sampling. Again, we find $N^\mathrm{MOR}_\omega=8$ yields perfect agreement with full CCGF for low-energy quasiparticle peaks.

\section{CCGF+\GW~tests on ZnO}
We performed detailed tests on the accuracy of active-space self-energy correction scheme on wurtzite ZnO using def2-SV(P) basis set and $2\times2\times1$ $\veck$-mesh. Band gaps of ZnO from MOR-CCGF+\GW~calculations are summarized in Table~\ref{tab:znocas}. We find that by increasing the size of active space, CCGF+\GW~band gaps gradually approach the full CCGF value (i.e., 2.80 eV). In the meantime, one needs to use relative large active space, such as (36e, 36o) per unit cell, to reach under 5\% relative error.

\begin{table}[hbt]
	\centering
	\caption{Band gaps of wurtzite ZnO computed by MOR-CCGF+\GW. }
	\label{tab:znocas}
	\begin{tabular}{>{\centering\arraybackslash}p{4cm}>{\centering\arraybackslash}p{2cm}>{\centering\arraybackslash}p{3cm}}
	\hline\hline
	 Active Space &  Band Gap (eV) &  Relative Error (\%)  \\
	\hline
	(16e, 18o) & 3.02 & 7.8 \\
	(16e, 34o) & 2.96 & 5.9 \\
    (20e, 20o) & 3.01 & 7.8 \\
    (28e, 28o) & 3.00 & 7.3 \\
    (36e, 36o) & 2.92 & 4.3 \\
    (44e, 44o) & 2.89 & 3.3 \\
    (52e, 52o) & 2.88 & 2.9 \\
    (60e, 60o) & 2.82 & 0.7 \\
    (76e, 76o) & 2.80 & 0.0 \\
    \hline\hline
	\end{tabular}
\end{table}

\begin{figure*}[htb!]
    \centering
     \begin{subfigure}
         \centering
         \includegraphics[width=0.4\textwidth]{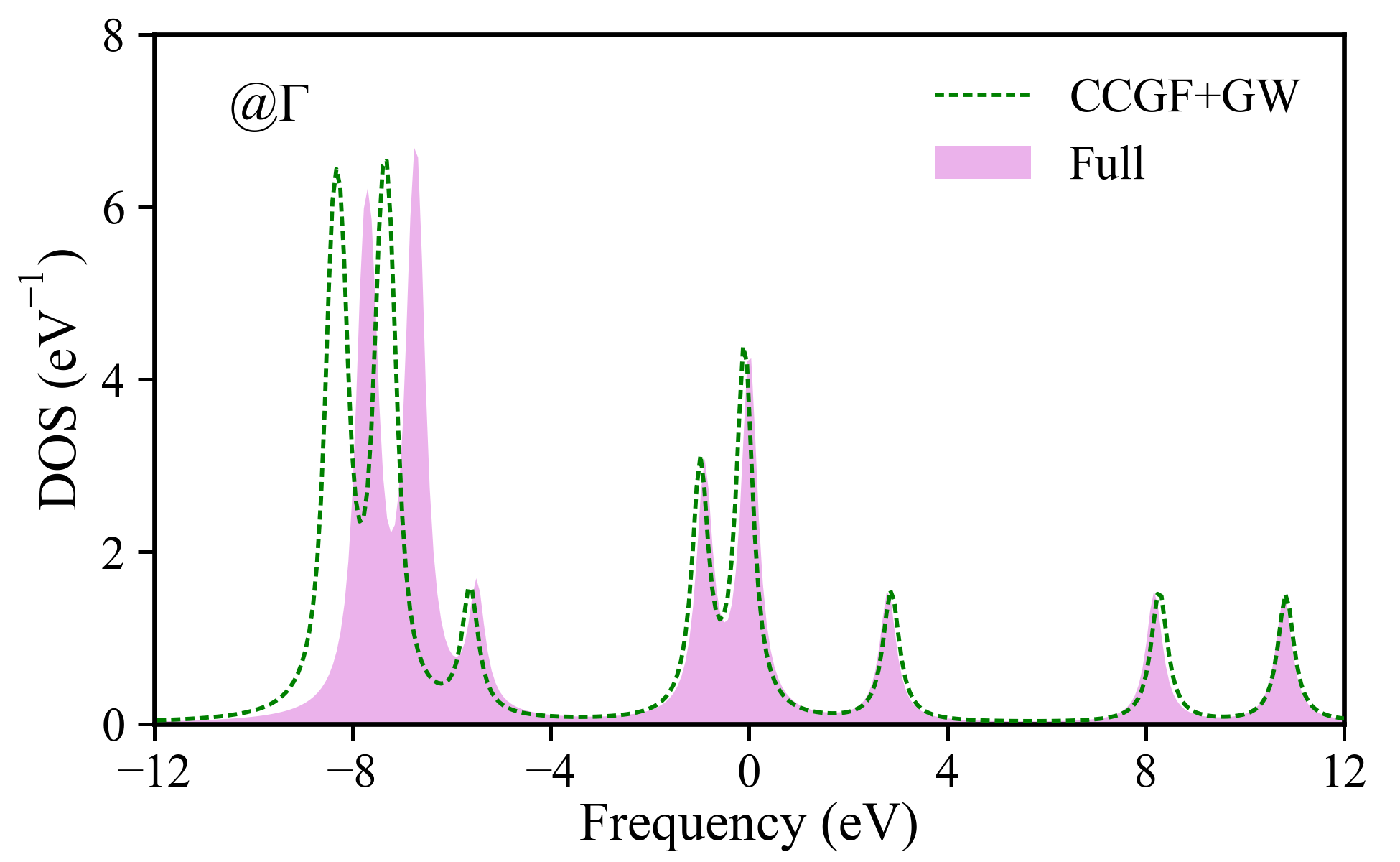}
     \end{subfigure}
     \hspace{1em}%
     \begin{subfigure}
         \centering
         \includegraphics[width=0.4\textwidth]{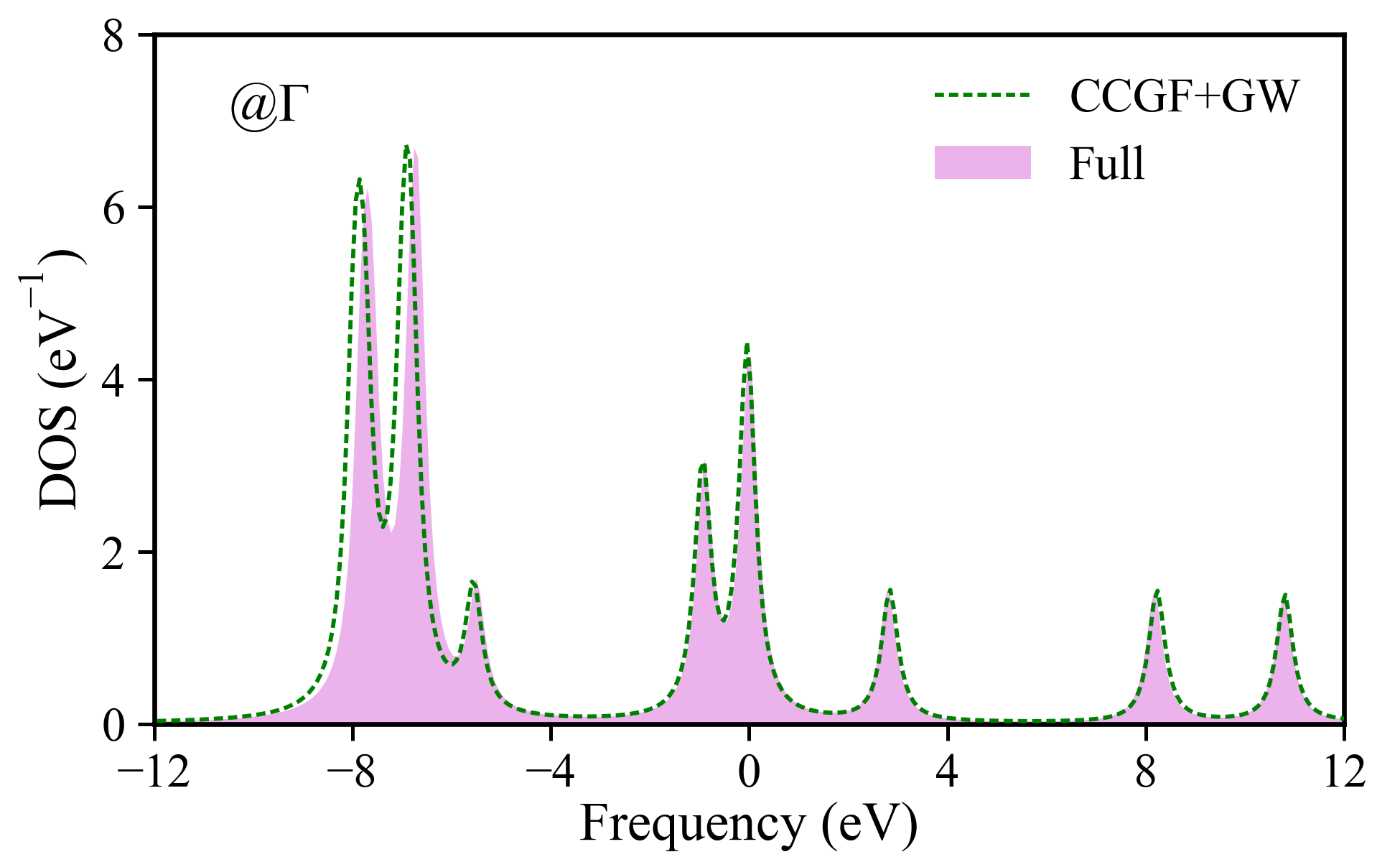}
     \end{subfigure}
    \caption{DOS of ZnO at $\Gamma$ point computed by MOR-CCGF+\GW. $N^\mathrm{MOR}_\omega=8$ and $\eta=0.2$ eV. Left: DOS using (16e, 34o) active space per unit cell. Right: DOS using (52e, 52o) active space per unit cell.} 
    \label{fig:znocas}
\end{figure*}

In Fig.~\ref{fig:znocas}, we show DOS of ZnO at $\Gamma$ point computed by MOR-CCGF+\GW~at two different active spaces. We find that, compared to (16e, 18o) MOR-CCGF+\GW~calculation in Fig.~\ref{fig:CAS}(d), (16e, 34o) MOR-CCGF+\GW~only improves the description of low-energy bands slightly, while the discrepancy in the [-10, -6] eV region remains. When the active space is increased to (52e, 52o), we finally find good agreement with the full MOR-CCGF DOS over all frequency regions.

\end{document}